\documentclass[fleqn,usenatbib]{mnras}

\usepackage{newtxtext}

\usepackage[T1]{fontenc}

\DeclareRobustCommand{\VAN}[3]{#2}
\let\VANthebibliography\thebibliography
\def\thebibliography{\DeclareRobustCommand{\VAN}[3]{##3}\VANthebibliography}

\usepackage{graphicx}	
\usepackage{amsmath}	
\usepackage{amssymb}	

\newcommand{\deriv}[2]{\ensuremath{\frac{{\rm d} #1}{{\rm d} #2}}}
\newcommand{\pderiv}[2]{\ensuremath{\frac{\partial #1}{\partial #2}}}

\title[Dust delivery to ionized winds]{Modeling the delivery of dust from discs to ionized winds}

\author[Booth \& Clarke]{
Richard A. Booth,$^{1}$\thanks{E-mail: r.booth@imperial.ac.uk}
and Cathie J. Clarke,$^{2}$\\
$^{1}$Astrophysics Group, Imperial College London, Prince Consort Road, London SW7 2AZ, UK\\
$^{2}$Institute of Astronomy, Madingley Road, Cambridge CB3 0HA, UK\\
}

\date{Accepted 2021 January 11. Received 2021 January 11; in original form 2020 December 10}

\pubyear{2020}

\begin{document}
\label{firstpage}
\pagerange{\pageref{firstpage}--\pageref{lastpage}}
\maketitle

\begin{abstract}
A necessary first step for dust removal in protoplanetary disc winds is the delivery of dust from the disc to the wind. In the case of ionized winds, the disc and wind are sharply delineated by a narrow ionization front where the gas density and temperature vary by more than an order of magnitude. Using a novel method that is able to model the transport of dust across the ionization front in the presence of disc turbulence, we revisit the problem of dust delivery. Our results show that the delivery of dust to the wind is determined by the vertical gas flow through the disc induced by 
the mass loss, rather than turbulent diffusion (unless the turbulence is strong, i.e. $\alpha \gtrsim 0.01$). Using these results we provide a simple relation between the maximum size of particle that can be delivered to the wind and the local mass-loss rate per unit area from the wind. This relation is independent of the physical origin of the wind and predicts  typical sizes in the 0.01 -- $1\,\micron$ range for EUV or X-ray driven winds. These values are a factor $\sim 10$ smaller than those obtained when considering only whether the wind is able to carry away the grains.
\end{abstract}




\section{Introduction}

Mass-loss in protoplanetary disc through winds is important for understanding their evolution. In particular, photoevaporative winds driven by either X-rays, extreme, or far ultra-violet (EUV, FUV) radiation are thought to be responsible for the final rapid clearing of protoplanetary discs \citep[e.g.][]{Clarke2001,Owen2011b,Ercolano2015,Gorti2016}. More recently, magneto-hydrodyanmic (MHD) winds have replaced turbulence as being the most promising processes responsible for driving  accretion in protoplanetary discs \citep{Salmeron2007,Suzuki2009,Bai2017,Bethune2017}. As a result, the entrainment of dust in these winds has also become an important issue.  \citet{Throop2005} suggested that the preferential removal of gas by winds might aid planet formation although more recent studies \citep{Ercolano2017,Sellek2020} suggest that this is unlikely unless radial drift of dust can be suppressed. In addition, the  dust entrained in  winds may provide a way to probe them observationally \citep[e.g.][]{Owen2011a, Miotello2012,Franz2020}.

The problem of dust entrainment may be thought of in two parts. Firstly, there is the question of whether dust particles entering the wind region are sufficiently well coupled to the gas so as to  be carried away by the wind. However, more importantly, such escaping  dust grains also need to  be {\it delivered}  to the wind from the underlying disc. In the case of winds driven by EUV radiation, the wind and disc are sharply delineated by a narrow ionization front where the gas density and temperature vary by many orders of magnitude.  Previously,  \citet[]{Hutchison2016, HC20} have found that delivery is the limiting step in controlling the range of dust sizes that are lost in such winds.

 In some previous studies, \citep[e.g.][]{Hutchison2016} it has been assumed that dust delivery to the wind  occurs diffusively, with turbulence in the disc competing against settling due to gravity to loft the grains into the wind. However, it has recently become clear that dust may instead be delivered to the wind \emph{advectively}, through coupling of the motion of dust grains to the upward  motion  of gas towards the ionization front. 
 Although this gas motion is strongly subsonic (likely even below the typical turbulent speeds), \citet{HC20} argued that advection is an important component in the delivery of dust to the wind. There is precedence for this, since in their MHD simulations of disc winds \citet{Riols2018} showed that advection resulted in an increase of the dust scale height over the height expected from purely turbulent transport. 

However, \citet{HC20} encountered a problem 
in quantifying  how efficiently dust is delivered to the wind. This problem was associated with the non-convergence of their results as the width of the ionization front was reduced.  This non-convergence, as discussed in \citet{HC20}, results from the fact that  there is a steep gradient in  gas density at the ionization front, where the gas goes from being cold in the disc to hot in the ionized wind. Although the gas velocity changes rapidly across the ionization front, changes in the dust velocity occur on a `stopping time', $t_{\rm s}$, the time over which drag forces act. This means that  the dust density varies  over the `stopping length', $v t_{\rm s}$. Since the stopping length can be much larger than the ionization front width, this leads to a steep increase of  the dust-to-gas ratio across the ionization front.  \citet{HC20} modelled the effects of turbulence as a diffusion equation \citep[following][]{Dubrulle1995} so that this  large gradient in dust-to-gas ratio produced a  large negative  diffusive flux.  This flux  then suppresses the delivery of dust to the wind, by an amount that depends on the width of the ionization front.

 However, this approach is not fully consistent because the diffusion is ultimately driven by the coupling of the dust dynamics to  turbulent gas motions via  drag forces. Diffusive motions are therefore subject to the same constraints as the mean flow in being limited by the finite coupling between dust and gas.   A reduction of the  effective difffusion coefficient, in cases where the gas flow changes on scales less than the stopping distance,  is however not captured by the formulation of \citet{Dubrulle1995}, which therefore gives erroneous results in this limit.


The goal of this paper is to rectify this deficiency in modeling  dust transport across narrow fronts, thus determining how efficiently dust is delivered to the wind. Instead  of solving an advection-diffusion equation, we  use a Monte-Carlo model to trace the  dynamics of individual dust grains, explicitly treating the coupling of the dust to turbulent velocity fluctuations in the disc gas. This approach is similar to the one used by \citet{Youdin2007} to model the diffusion of large dust particles in discs, for example. We present our model in \autoref{sec:model}, and in \autoref{sec:tests} demonstrate that, in contrast to other formulations for modeling dust in turbulent flows in the literature, we are able to correctly recover the structure of the gas and dust across steep transitions in the gas density, an important prerequisite for tackling problems involving ionization fronts. In \autoref{sec:analytic} we 
consider the two  criteria suggested by \citet{HC20} as limiting the maximum size of grains that are a) deliverable to the ionization front and b) entrainable by the wind above the ionization front, for which the corresponding Stokes numbers (evaluated just below the ionization front) are denoted by  $St_{\rm{crit}}$ and $St_{\rm{max}}$ respectively). We then  use these limits to estimate the level of turbulence at which a  transition between diffusive and advective feeding of dust into the wind base  is expected. In \autoref{sec:results} we demonstrate that,
as anticipated by \citet{HC20}, 
$St_{\rm{crit}}$ (which turns out to be $\sim 0.01 $  for a wide range of input parameters) indeed represents a good limit for setting the maximum size of dust delivered into the wind: although somewhat larger dust grains may enter the 
wind in the limit of strong turbulence, the grains entering the wind have Stokes number significantly less than $St_{\rm{max}}$ and hence are all capable of being fully entrained in the ionized flow. A discussion of our results and conclusions are presented in \autoref{sec:discussion} and \autoref{sec:conclusions}.

\section{Model}
\label{sec:model} 
\subsection{Modeling the disc/wind base: gas}
\label{sec:disc_model}

We model the entrainment of dust in a background disc undergoing photoevaporation. The vertical structure of the disc is computed by solving the momentum equation of hydrodynamics in one dimension,
\begin{equation}
    u_{\rm z} \pderiv{u_{\rm z}}{z} = - \frac{1}{\rho} \pderiv{\left[\rho c_s(z)^2\right]}{z} -  \frac{G M z}{(R^2 + z^2)^{3/2}}, \label{eqn:mom_eqn}
\end{equation}
assuming steady-state such that $\rho u_{\rm z}$ is constant. Here $u_{\rm z}$ is the gas velocity, and $\rho$ is the gas density. 

The sound-speed profile, $c_{\rm s}(z)$, is chosen to model the transition from a cold disc to a hot photoionized wind at the ionization front, 
\begin{equation}
    c_{\rm s}(z) = \frac{c_{\rm s, disc} + c_{\rm s,wind}}{2} + \frac{c_{\rm s, wind} - c_{\rm s,disc}}{2} \tanh\left(\frac{|z - z_{\rm IF}|}{3W}\right). \label{eqn:c_s}
\end{equation}
Here disc is assumed to be vertically isothermal, where $c_{\rm s, disc}$ and $c_{\rm s,wind}$ denote the sound speed in the disc and wind, with $z_{\rm IF}$ and $W$ specifying the location and width of the transition. 

By default we take $M=1\,M_\odot$ and $R=10\,{\rm au}$. We assume the disc aspect ratio is given by $H/R = 0.05 (R/{\rm au})^{0.25}$, thus ${c_{\rm s, mid} \approx 0.84\,{\rm km\,s}^{-1}}$. The sound speed in the wind, ${c_{\rm s,wind} = 12.85\,{\rm km\,s}^{-1}}$, appropriate for a fully ionized hydrogen gas at $10^4 {\rm K}$. Where required, the mid-plane density is taken to be $\rho_0= \Sigma(R) / {\sqrt{( 2 \upi)}} H$, assuming the gas surface density $\Sigma(R) = 30 (R/{\rm au})^{-1}\,{\rm g\,cm}^{-3}$.

To set the mid-plane velocity, we assume that photoionization drives an outflow with a velocity of $v_{\rm wind} = 0.5 c_{\rm s, wind}$ at the ionization front, this being motivated by typical launch velocities for self-similar solutions for isothermal winds \citep{Clarke2016}. Explicitly, we find the velocity at $z=0$ iteratively by integrating \autoref{eqn:mom_eqn} to $z = z_{\rm IF} + 9W$ and requiring that $v_z$ at this point is $0.5 c_{\rm s, wind}$. The parameter $z_{\rm IF}$ controls the density at the base of the ionized wind $\rho_{\rm ion}$, once the mid-plane density and temperature of the disc are assigned. Since the disc below the ionization front is very close to a state of hydrostatic equilibrium, $\rho_{\rm ion} \sim  \rho_0 \exp(-z_{\rm IF}^{2}/2H^2) f_i$, where $f_i$ is the factor by which the density drops across the ionization front: $f_i \sim  c_{\rm s, mid}^2/(c_{\rm s, wind}^2 + v_{\rm wind}^2)$. The canonical
parameters detailed above and $\rho_{\rm ion}$ derived from \autoref{eqn:rho_ion} for an ionizing flux of $10^{42}\,{\rm s}^{-1}$ corrsponds to $z_{IF} \sim 4H$. Assuming the standard profiles for photoevaporative winds driven by EUV radiation \citep{Hollenbach1994}, these values  would correspond to integrated mass loss rates from the disc of order $10^{-10}-10^{-9} M_\odot {\rm yr}^{-1}$.

\autoref{eqn:mom_eqn} is solved numerically using the 4th-order Runge-Kutta method of Dormand \& Prince \citep[e.g.][]{PressNR} as implemented in the \textsc{odeint} package in the \textsc{boost} library\footnote{\href{https://www.boost.org/}{https://www.boost.org/}}. The solution at intermediate points is then obtained via piecewise-cubic Hermite interpolation \citep{PCHIP}.

\subsection{Modeling the disc/wind base: dust}
\label{sec:dust_model}

The dust component is treated using a Stochastic Lagrangian Model. We compute the trajectory of a large number of tracer particles under the action of gravity, coupled to the gas via drag forces:
\begin{align}
    \deriv{z}{t} &= v_{\rm z}, \\ 
    \deriv{v_{\rm z}}{t} &=  - \frac{v_{\rm z} - [u_{\rm z}(z) + u'_{\rm z}(z,t)]}{t_{\rm stop}}  - \frac{G M z}{(R^2 + z^2)^{3/2}}, \label{eqn:dust_accel}
\end{align}
where $v_{\rm z}$ is the vertical velocity of the dust. Here we have decomposed the gas velocity into its background component and a fluctuating part, $u'_{\rm z}(z,t)$, which represents the motions due to turbulence in the disc that are responsible for diffusion. We will assume that the turbulent fluctuations are Gaussian in nature with a correlation time, $t_{\rm e}$. We assume linear Epstein drag such that,
\begin{equation}
    t_{\rm stop} = \sqrt{\frac{\upi}{8}} \frac{\rho_{\rm grain} s}{c_s \rho}
\end{equation}
where $\rho_{\rm grain}$ is the internal density of a dust grain and $s$ is its size. 
Typically, we label grain size by their Stokes number, $St = t_s \Omega$, but where relevant we will assume $\rho_{\rm grain} = 1\,{\rm g\,cm}^{-3}$. The background gas velocity, $u_{\rm z}(z)$, sound speed, $c_s$, and density, $\rho$, are taken from the model described in \autoref{sec:disc_model}.

The turbulent fluctuations are treated using a Langevin model based on \citet{Thomson1984}  (see also \citealt{Wilson1983}). \citet{Thomson1984} derived a Stochastic Lagrangian model for the motion of tracer particles in the atmosphere by requiring that the statistical distribution of the particles must be the same as that of the underlying atmosphere. \citet{Thomson1987} showed that this requirement -- that the particles must remain `well mixed' with the gas -- is rather general, with Stochastic Lagrangian models that satisfy this criterion being consistent with the Euler equations and able to reproduce both the short and long term behaviour of the gas. Under the assumption of Gaussian velocity fluctuations, the model for the gas is
\begin{align}
\delta z &= [u_{\rm z}(z) + \sigma(z) w_{\rm t}] \delta t, \\
\delta w_{\rm t}  &= - \frac{(w_{\rm t} - w_{\rm hs})}{t_{\rm e}} \delta t  + \sqrt{\frac{2}{t_e}} \delta W \label{eqn:w_t},
\end{align}
where
\begin{equation}
w_{\rm hs} = \sigma(z) \left\{ \frac{1}{2} \pderiv{\ln [\sigma(z)^2]}{z} + \pderiv{ \ln [\rho(z)]}{z}  \right\} t_{\rm e}.
\end{equation}
The $w_{\rm hs}$ term corrects for the fact that that a Gaussian distribution of turbulent velocities with mean zero will drive a non-zero net flux when there are gradients in $\rho(z)$ or $\sigma(z)$ \citep{Thomson1984, Ciesla2010}. Here $t_{\rm e}$ is the Lagrangian correlation time, $\sigma(z,t)^2$ is the variance of the velocity fluctuations and $W$ is Wiener Process, i.e. $\delta W$ is a random number distributed as $\delta W \sim \mathcal{N}(0, \delta t)$ (where $\mathcal{N}(0, \delta t)$ is a normal distribution with mean zero and variance $\delta t$). The diffusion coefficient $D$ is linked to $t_{\rm e}$ and $\sigma$ via $D = \sigma^2 t_{\rm e}$ \citep[e.g.][]{Youdin2007,Ormel2018}. By default we take $t_{\rm e} = \Omega^{-1}$, where $\Omega$ is the Keplerian frequency and $D=\alpha c_{\rm s, disc}^2 \Omega^{-1}$, giving $\sigma = \sqrt\alpha c_{\rm s, disc}$ so that the dimensionless parameter $\alpha$ relates the sound speed to the turbulent velocity as in the viscous $\alpha$ parametrisation of \citet{Shakura1973}.

We extend this model to treat dust grains in the simplest way possible, which is to use take $u_{\rm z}'(z) = \sigma(z, t) w_{\rm t}$ and use the particle's position to define $z$ in \autoref{eqn:w_t}. Explicitly, we use
\begin{align}
    \delta z &= v_{\rm z} \delta t, \label{eqn:z_dust} \\
    \delta v_{\rm z} &=  - \frac{v_{\rm z} - [u_{\rm z}(z) + \sigma(z) w_{\rm t}]}{t_{\rm stop}} \delta t - \frac{G M z}{(R^2 + z^2)^{3/2}} \delta t, \label{eqn:dust_eom} \\
    \delta w_{\rm t}  &= - \frac{(w_{\rm t} - w_{\rm hs})}{t_{\rm e}} \delta t  + \sqrt{\frac{2}{t_e}} \delta W, \label{eqn:w_turb}
\end{align}
which reduces to \citet{Thomson1984}'s model in the limit ${t_{\rm stop} \rightarrow 0}$. 

Our model is similar to, but differs from, existing Stochastic Lagrangian Models for dust in the literature. For $u_{\rm z} = 0$ and constant $\rho$, our model reduces to that of \citet{Youdin2007}. The model of \citet{Ormel2018} is the most similar to ours, differing  by the way in which the correction term $w_{\rm hs}$ is implemented. \citet{Ormel2018} add a term $\sigma w_{\rm hs}$ to $u_{\rm z}$ in \autoref{eqn:dust_eom} while neglecting $w_{\rm hs}$ in \autoref{eqn:w_turb}. When $w_{\rm hs}$ is slowly varying the effect of this on the dynamics is small; however, in the presence of a sharp transition in the density (or turbulence), as is the case at an ionization front, the difference becomes significant. Since we apply the correction in \autoref{eqn:w_turb} the effects of steep transition in density are averaged over $t_{\rm e}$, whereas in the case of \citet{Ormel2018} they are applied locally. The model proposed by \citet{Laibe2020} is equivalent to assuming $w_{\rm hs} = 0$ in \autoref{eqn:w_turb}. These differences are highlighted in \autoref{sec:tests}\footnote{\citet{Ciesla2010} also provided a Stochastic Lagrangian Model that satisfies \citet{Thomson1987}'s well-mixed condition. However, \citet{Ciesla2010} used the terminal velocity approximation for the mean flow and, by imposing fixed (i.e. $t_{\rm stop}$ independent) velocity impulses,  neglected the finite Lagrangian correlation time of the turbulence, making it inappropriate for the ionization front problem. Thus we do not consider it further.}. 

\subsection{Implementation}

We have adopted a semi-implicit approach to solve equations~\ref{eqn:z_dust}--\ref{eqn:w_turb} efficiently for particles with short stopping times. First, we move the particle from $z$ to $z + v_{\rm z} \delta t$. Next we update $w_{\rm t}$, evaluating the right-hand side of \autoref{eqn:w_turb} at the new position. Finally, we update the dust velocity, $v_{\rm z}$, using the new position and $w_{\rm t}$. This update is done implicitly in $v_{\rm z}$ to avoid limits on the time-step due to small $t_{\rm stop}$, such that
\begin{align} 
    v_{\rm z} (t + \delta t) = \left[v_{\rm z}(t) - \frac{G M z \delta t}{(R^2 + z^2)^{3/2}}\right] &\frac{t_{\rm stop}}{\delta t + t_{\rm stop}} \nonumber \\ 
    \qquad + \left[u_{\rm z}(z) + \sigma(z)w_{\rm t}\right] & \frac{\delta t}{\delta t + t_{\rm stop}}.
\end{align}
It is straightforward to verify that this expression is correct in the limits $t_{\rm stop} \rightarrow 0$ and $t_{\rm stop} \rightarrow \infty$.

The time-step, $\delta t$, is chosen to satisfy a number of constraints:
\begin{align}
   \delta t_{\rm max} &= \left[ \delta t_0^{-1} + \delta t_1^{-1} + \delta t_2^{-1} + \delta t_3^{-1}  \right]^{-1} \nonumber \\
   \delta t_0 &= 0.01 t_{\rm e}  \nonumber \\
   \delta t_1 &= 0.01\left(\pderiv{\sigma}{z}\right)^{-1} \nonumber \\
   \delta t_2 &= 0.01\left(\sigma \pderiv{\ln \rho}{z}\right)^{-1} \nonumber \\
   \delta t_3 &= 0.05 \frac{ \max(|z - z_{\rm IF}|, W)}{v_{\rm z}}. \nonumber 
\end{align}
The first three constraints are designed to ensure that $w_{\rm t}$ and $w_{\rm hs}$ do not change significantly in one time-step, while the last one is included to make sure that the particles do not jump across the ionization front in a single time-step.

Rather than setting $\delta t = \delta t_{\rm max}$, we instead set $\delta t =  2^{\ell-\ell_{\rm max}}\Omega^{-1}$, where $\ell_{\rm max} = 63$ and $\ell$ is the largest integer such that $\delta t \le \delta t_{\rm max}$. Particle time steps are always allowed to decrease; however, increases are only allowed if the particle would remain synchronised (i.e. $t$ is exactly divisible by $\delta t$). This decision is made to ensure that all particles are synchronised every $t_{\rm e}$, so that their positions and velocities can be sampled at the same time.

\section{Tests}
\label{sec:tests}

\begin{figure*}
	\includegraphics[width=\textwidth]{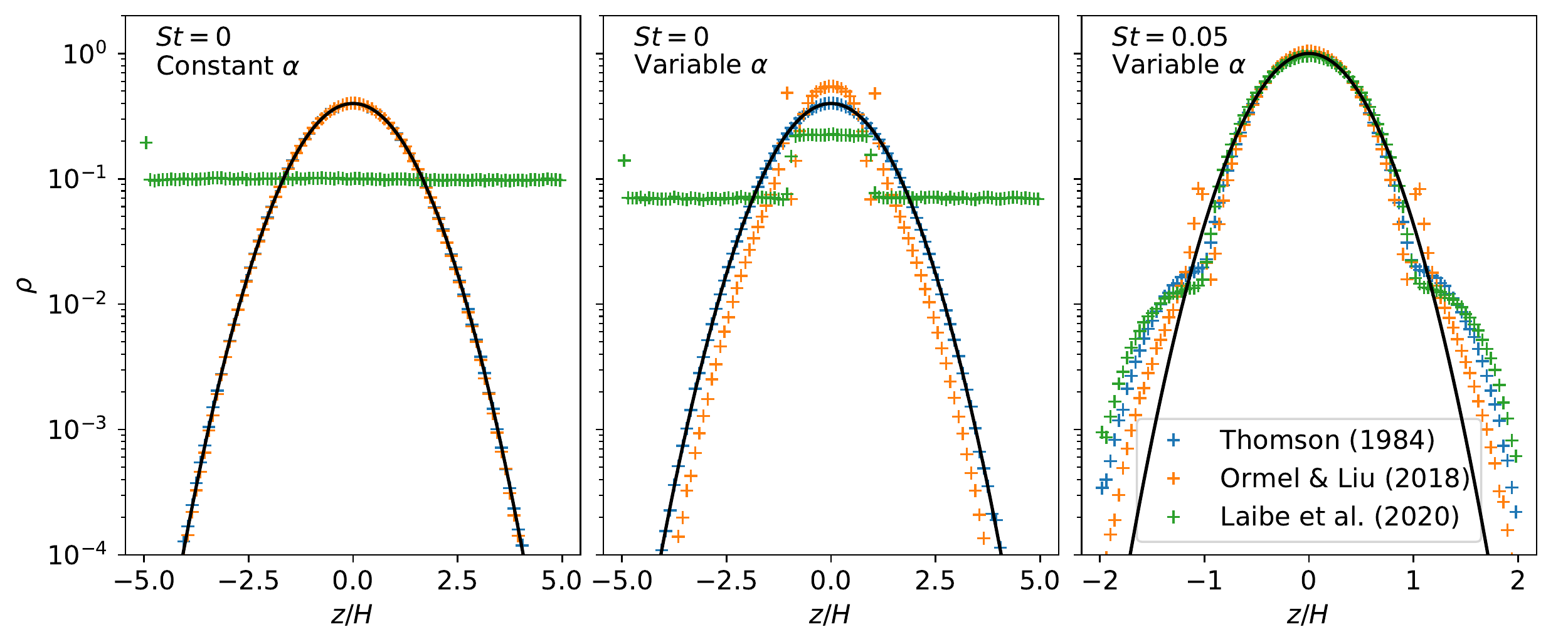}
    \caption{The vertical structure of gas and dust particles in a Gaussian disc computed with different Stochastic Lagrangian Models. Left: a constant turbulent $\alpha=0.01$ is used. Middle and right: $\alpha$ transitions from 0.01 to 0.1 at $z=H$. In each case the black line shows the analytical solution (only valid close to the mid-plane for $St = 0.05$). Only the \citet{Thomson1984} model (which we use in this paper) recovers a constant dust-to-gas ratio for $St=0$ in the presence of steep gradients.}
    \label{fig:tests}
\end{figure*}

To test the code, we compute the distribution of particles in a Gaussian disc with a constant sound speed and $u_{\rm z} = 0$, comparing the results to the methods proposed by \citet{Ormel2018} and \citet{Laibe2020}. For each test, $10^3$ particles were injected at z=0. After a burn-in period of $10^4\Omega^{-1}$ the positions of each particle in the range $[-5,5]$ (for $St=0$ and $[-2,2]$ for $St=0.05$) were recorded every $10\Omega^{-1}$ for the next $10^5\Omega^{-1}$. The density was computed by binning the particles into 100 bins, normalised such that the total mass is 1.

First we consider particles with $St=0$, which should be distributed with the same density as the gas. \autoref{fig:tests} (left panel) shows this for the case of a constant $\alpha = 0.01$. Here both our method and \citet{Ormel2018}\footnote{Note that we updated the dust velocity implicitly, as in our method, rather than using \citet{Ormel2018}'s `Strong Coupling Approximation' since that method is not appropriate for our problem.} produce similar results, with the particles well-mixed with the gas. However, the method of \citet{Laibe2020} produces a constant dust density rather than dust-to-gas ratio, which is a consequence of neglecting the $w_{\rm hs}$ term. 

Next, we consider the same model but with $\alpha$ varying from 0.01 to 0.1, using the functional form of \autoref{eqn:c_s} with $z_{\rm IF}=H$ and $W=H/60$. Note that $c_s$ is constant in this test. Under these conditions  particles with $St=0$ should still have the same density distribution as the gas. Now we see a difference between the method of \citet{Ormel2018} and the one presented in this work, with only our method producing a constant dust-to-gas ratio. The `blip' in the density around $z=H$ produced by \citet{Ormel2018}'s method arises because $w_{\rm hs}$ is large at  this location. When applied in \autoref{eqn:dust_eom}, as in \citet{Ormel2018}, this leads to large velocities for the dust particles, which are responsible for the `blip'. In our method, these large velocities are not produced because $w_{\rm hs}$ gets averaged over $t_e$. We note that these differences are only significant if $w_{\rm hs}$ varies over a small length scale -- this was not the case in the tests presented in \citet{Ormel2018}, but such variations do occur close to the ionization front in our model. Again, the method of \citet{Laibe2020} produces constant dust density in regions where $\alpha$ is constant.

The right panel of \autoref{fig:tests} shows a repeat of the test with $\alpha$ varying with height for particles with $St=0.05$ at $z=0$. In this case all of the methods produce similar results close to $z=0$, which are in good agreement with the analytical solution of \citet[away from the mid-plane the analytical solution is no longer valid]{Youdin2007}. Again the \citet{Ormel2018} method shows an artefact at the transition, while in this case settling reduces the difference between  our method and that of \citet{Laibe2020}.

\begin{figure*}
	\includegraphics[width=\textwidth]{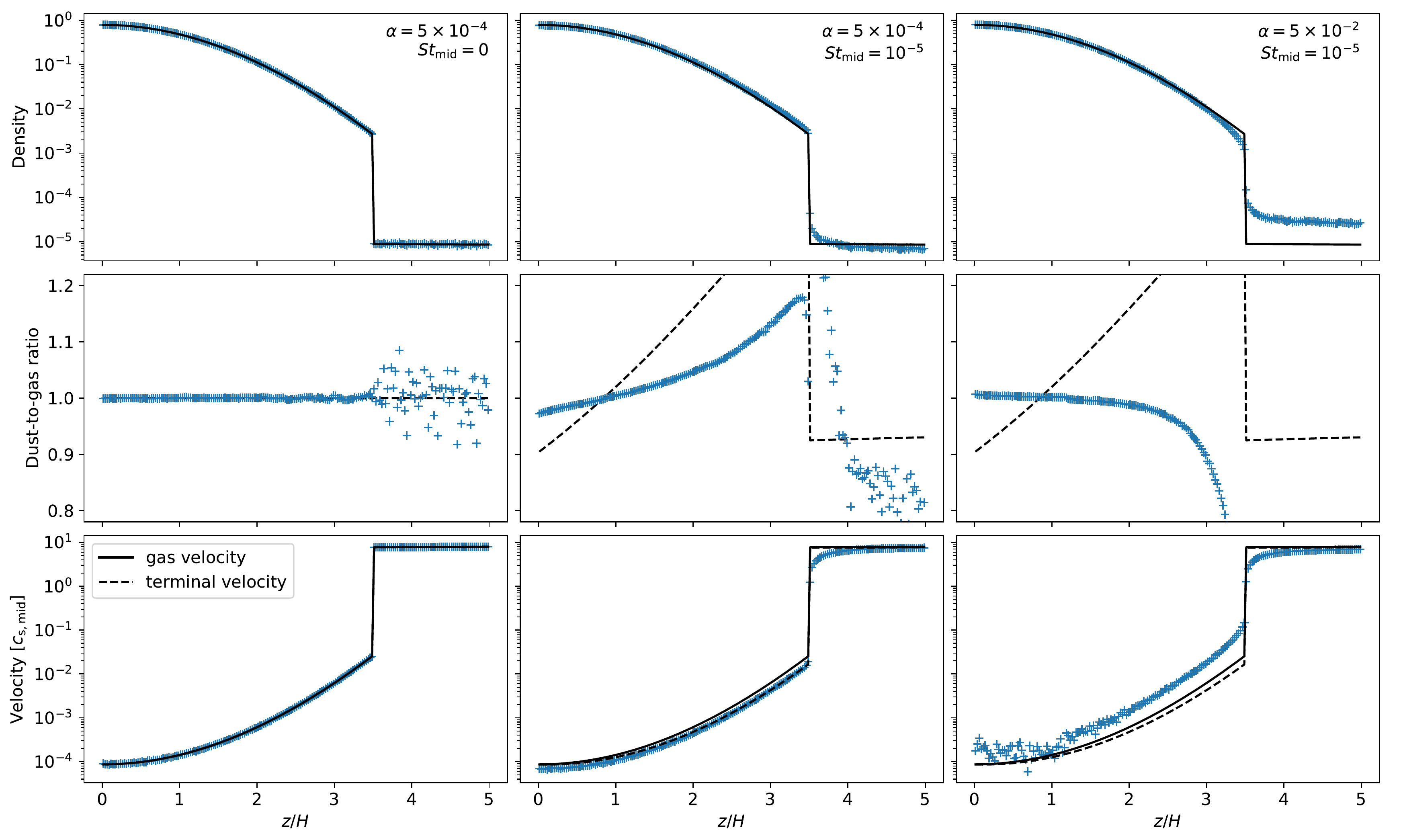}
    \caption{Vertical profiles of the density, dust-to-gas ratio and mean velocity as a function of height for three representative simulations (columns). In each panel the black solid lines show the gas properties from the disc model. The black dashed lines show the results expected for a purely advective solution in the terminal velocity limit.}
    \label{fig:profiles}
\end{figure*}

As a final confirmation of the ability of our code to deal with sharp gradients in the density we show in \autoref{fig:profiles} the density of $St=0$ particles in a ionization front test with a width, $W=10^{-5}\,{\rm au}$ (and  $H\approx 0.89\,{\rm au}$). This shows excellent agreement with the background profile, as expected.

\section{Analytic Estimates}
\label{sec:analytic}

Here we provide some simple estimates of the maximum size of dust grains that can be entrained in the wind. For sufficiently weak turbulence, the delivery of dust to the ionization front can be estimated by neglecting the $u'_{\rm z}$ term in \autoref{eqn:dust_accel}: in this case the passage of dust from the disc to the wind is simply set by the extent to which grains can couple to the advective flow of gas in the disc induced in response to mass loss at the ionization front. The maximum size of particles delivered to the wind may be estimated as the biggest particle for which $v_{\rm z} > 0$ at the ionization front, which, in the terminal velocity limit may be written as 
\begin{equation}
    St_{\rm crit} = \frac{u_{\rm z}(z_{\rm IF})}{\Omega z} \left(1 + \frac{z_{\rm IF}^2}{R^2} \right)^{3/2},
\end{equation}
as suggested by \citet{HC20}. Note that this refers to the Stokes number measured 
just below the ionization front. Since $u_{\rm z} \ll c_s$  in the disc, $St_{\rm crit}$ can be approximated as:
\begin{equation}
    St_{\rm crit} \approx \frac{\mathcal{M}_{\rm w}}{1+\mathcal{M}_{\rm w}^2} \frac{c_{\rm s,disc}}{c_{\rm s, wind}}\frac{H}{z_{\rm IF}} = 0.4 \frac{c_{\rm s,disc}}{c_{\rm s, wind}}\frac{H}{z_{\rm IF}},
    \label{eqn:stcrit}
\end{equation}
where we have used the Rankine-Hugoniot relations to relate the velocity in the disc to the Mach number at the base of the wind, ${\mathcal{M}_{\rm w} = 0.5}$. For typical values of $c_{\rm s, disc}$ and $c_{\rm s, wind}$, we find ${St_{\rm crit} \approx 0.01}$.

For comparison, the maximum particle size that, once in the wind, can escape from the disc is approximately given by the particle size for which the terminal velocity is zero at the base of the wind, i.e. downwind of the ionization front  \footnote{Note that the condition that the terminal velocity is zero is equivalent to requiring a situation of zero net acceleration on a stationary particle as argued by  \citet{Takeuchi2005} and, with some order unity corrections for the flow geometry, also by \citet{HC20}}. Following, \citet{HC20}, we refer to this as $St_{\rm max}$, which is given by
\begin{equation}
    St_{\rm max} = (1+\mathcal{M}_{\rm w}^2) \frac{c_{\rm s,wind}}{c_{\rm s, disc}} St_{\rm crit} \approx  \mathcal{M}_{\rm w} \frac{H}{z_{\rm IF}} = 0.5 \frac{H}{z_{\rm IF}}.
    \label{eqn:stmax}
\end{equation}

We emphasise that although $St_{\rm crit}$ and $St_{\rm max}$ relate to situations of force balance applied on either side of the ionization front, they relate to stopping times that are evaluated at the same location (i.e. just below the ionization front) and their ratio therefore directly relates to the ratio of dust sizes that achieve this condition on each side of the front. Comparison of equations (\ref{eqn:stcrit}) and (\ref{eqn:stmax}) immediately shows that in the case of dust that is advected through an ionization front, the grains that are just able to reach the ionization front are a factor $\sim \frac{c_{\rm s,disc}}{c_{\rm s, wind}}$ in size below the maximum size that can be entrained in the ionized wind. Thus {\it delivery} of grains to the ionization front is the limiting step in removing dust from the disc rather than the subsequent ability of the ionized wind to carry it away \citep{HC20}.

When turbulence is strong, we expect that dust may be delivered to the ionization front diffusively instead of being delivered by advection. Neglecting the contribution from advection (i.e. setting $u(z) = 0$), the dust density is given by
\begin{equation}
    \frac{\rho_{\rm d}}{\rho_{\rm g}} \propto \exp\left( -\frac{St(z)}{\alpha}\right) 
\end{equation}
\citep{Dubrulle1995, Takeuchi2002}. Therefore, in a purely diffusive disc the delivery of dust to the ionization front should drop once ${St(z_{\rm IF}) > \alpha}$ (note we have neglected the influence of the ionization front itself). Since advection can efficiently supply dust to the ionization front for sizes below $St_{\rm crit}$, we therefore expect the transition to the diffusive regime to occur at ${\alpha \sim St_{\rm crit} \approx 0.01}$.

\begin{figure*}
	\includegraphics[width=\textwidth]{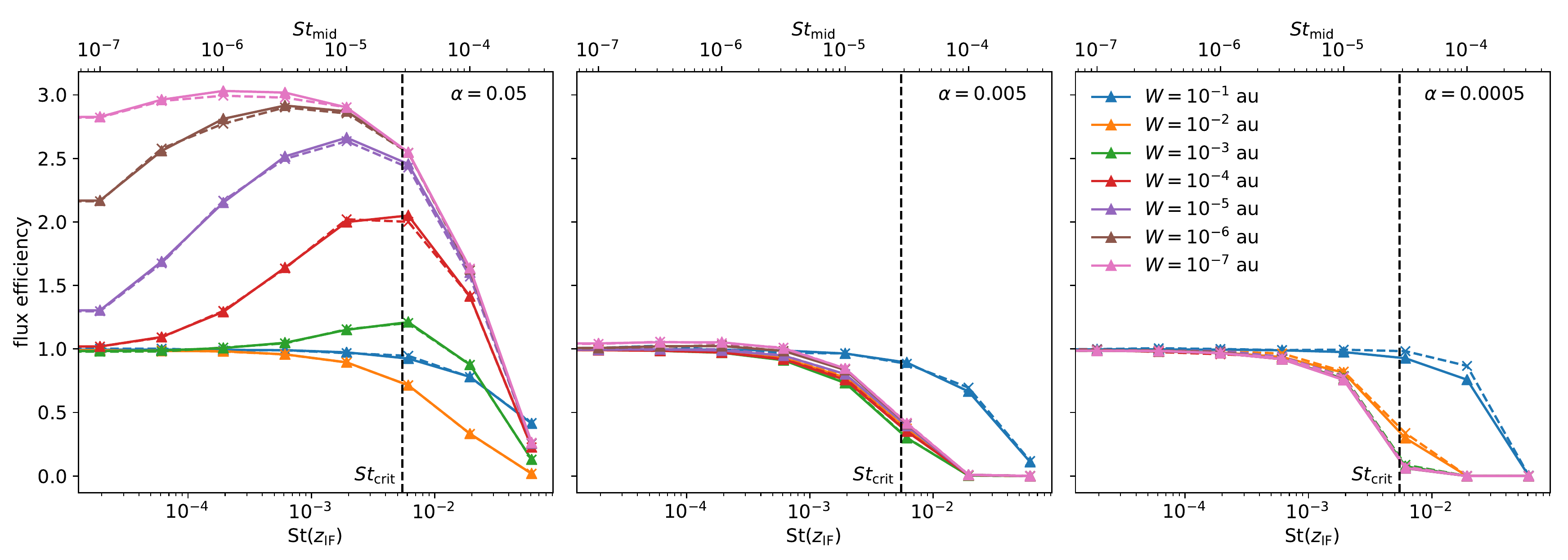}
    \caption{Flux efficiency, $\epsilon_{\rm F}$, for an ionization front model at $R=10\,\mathrm{au}$ and $z_{\rm IF}=3.5\,H$. Each panel shows a range of ionization front widths, $W$, for a different strength of turbulence, $\alpha$. For comparison, the flux efficiency computed using the definition in \citet{HC20} is shown by the dashed lines.}
    \label{fig:flux-efficiency}
\end{figure*}

\section{Numerical simulations}
\label{sec:results}

\subsection{Advective and diffusive dust delivery}

For the results presented in this section we set up the gas profile according to the description in \autoref{sec:disc_model}. Dust particles are then injected continuously at $z=0$ at a rate of $4\Omega$ and the simulation is run for $10^{5}\Omega^{-1}$. For the boundary conditions, particles are removed once they cross $z=5H$ while at $z=0$ we use reflecting boundaries (i.e. particles that cross $z=0$ have the sign of $z$, $v_{\rm z}$ and $w_{\rm t}$ flipped).

The mass-loss time-scale of dust particles was computed by comparing the rate of particle injection to the total number of particles in the domain once the simulation has reached a steady state. Since for very low mass-loss rates (i.e. for $St > St_{\rm crit}$) steady state is not reached within $10^5\Omega^{-1}$ we instead fit a model to the total number of particles in the domain over time using least-squares (via \textsc{scipy}'s \textsc{curve\_fit} routine\footnote{https://www.scipy.org/}). The model we use is
\begin{equation}
    \deriv{N}{t} = \dot{N}_0 - \frac{N}{\tau},
\end{equation}
where $\dot{N}_0 = 4 \Omega$. We then compare $\tau$ (i.e. $N/\dot{N}_0$ in steady-state) to the mass-loss time-scale of the gas $\Sigma/\dot{\Sigma}$ to determine the efficiency of dust entrainment in the wind,
\begin{equation}
    \epsilon_{\rm F} = \frac{\Sigma}{\dot{\Sigma}\tau} = \frac{\Sigma}{\Sigma_{\rm d}} \frac{\dot{\Sigma_{\rm d}}}{\dot{\Sigma}}.
\end{equation}
With this definition, the mass-loss rate of dust is simply the product of $\epsilon_{\rm F}$, the mass loss rate of gas and the dust to gas ratio. We note that this definition of the entrainment efficiency is slightly different to the definition used by \citet{HC20}, who used ${\dot{N}_0/(\rho_d(z=0) u_{\rm z}(z=0))}$ (where $\rho_d$ is normalised such that the total mass is 1). This choice is dictated by practicality: the definition used here is easier to determine accurately. However, the estimates only differ by a factor $({\Sigma_{\rm g} / \Sigma_{\rm d}) (\rho_{\rm d}(z=0)/\rho_{\rm g}(z=0))  \approx 1}$.

For the simulations presented in this section, we choose $z_{\rm IF}$ in the approximate range of $3H$ -- $4H$ as a compromise between realistic mass-loss rates and computational expediency. Note the mass-loss time-scale, $\Sigma/\dot{\Sigma}$, is (approximately) given by 
\begin{align}
\frac{\dot{\Sigma}}{\Sigma} &\approx \frac{1}{\sqrt{2\upi}} \exp\left(-\frac{z_{\rm IF}^2}{2 H^2}\right)  \frac{\mathcal{M}_{\rm w}}{1 + \mathcal{M}_{\rm w}^2} \frac{c_{\rm s,disc}}{c_{\rm s,wind}} \Omega \\
& \approx 1 \times 10^{-2} \exp\left(-\frac{z_{\rm IF}^2}{2 H^2}\right) \left(\frac{R}{10\,\rm au}\right)^{-1/4} \Omega.
\end{align}
Therefore $z_{\rm IF} = 4 H$ corresponds to a reasonable mass-loss time-scale of $10^6\,{\rm yr}$ at $R=10\,{\rm au}$.

The efficiency of dust entraiment for models with $R=10\,{\rm au}$ and $z_{\rm IF}=3.5\,H$ are shown in \autoref{fig:flux-efficiency} for $\alpha = 5 \times 10^{-4}$ to $5 \times 10^{-2}$ and a range of ionization front widths, $W$. 

For small $\alpha$, we find a flux efficiency $\epsilon_{\rm F} \approx 1$ for small $St$, transitioning to $\epsilon_{\rm F} \approx 0$ at  $St \approx St_{\rm crit}$ for all but the largest of ionization front widths. This is the expected result for advection-dominated delivery of dust to the wind. Comparing the $\alpha = 5 \times 10^{-4}$ results to those $\alpha = 5 \times 10^{-3}$ shows that increasing $\alpha$ mildly increases $\epsilon_{\rm F}$ for $St$ close to $St_{\rm crit}$, but in both cases $St_{\rm crit}$ remains a good estimator of the maximum dust size that can be entrained.

For $\alpha = 0.05$ the delivery of dust to the ionization front is now diffusion dominated according to our estimate in \autoref{sec:analytic} (since $\alpha > St_{\rm crit}$), resulting in different behaviour. For $St \lesssim St_{\rm crit}$, the mass-loss rate of dust is now \emph{higher} than that of the gas and also dependent on the width of the ionization front. We also find that for sufficiently narrow ionization fronts the dust flux eventually converges. Convergence occurs at progressively smaller ionization front widths as the Stokes number decreases. This convergence can be explained by considering the stopping distance at the two different sides of the ionization front:
\begin{align}
    l_{\rm disc} \approx& \frac{\mathcal{M}_{\rm w}}{1+\mathcal{M}_{\rm w}^2} \frac{c_{\rm s,disc}}{c_{\rm s, wind}} St H = 0.4 \frac{c_{\rm s,disc}}{c_{\rm s, wind}} St H,\\
    l_{\rm wind} \approx&  l_{\rm disc} (1+\mathcal{M}_{\rm w}^2) \frac{c_{\rm s,wind}}{c_{\rm s, disc}}  = 0.5 St H,
\end{align}
where in both cases the Stokes number is measured in the disc immediately before the ionization front. As the width of the ionization front is successively decreased, particles with a given Stokes number will first decouple on the down-wind side of the ionization front. Finally once $W \ll l_{\rm disc} \approx 0.03 St H$, the particles will cross the ionization front before being able to react, and thus the width of the ionization can no longer affect the flux. \autoref{fig:flux-efficiency} confirms this, with convergence by $W \approx 0.1 l_{\rm disc}$.

The dependence of the flux on $\alpha$ can be understood by looking at the dust-to-gas ratio profiles, shown for two examples in \autoref{fig:profiles}. At $\alpha = 5 \times 10^{-4}$ and $St_{\rm mid} = 10^{-5}$, the dust-to-gas ratio increases with $z$ between the mid-plane and the ionization front. This follows from  mass conservation and the fact that dust velocity is lower than the gas velocity since the gravitational acceleration is not nearly balanced by pressure, as in the case of  the gas, and finite $t_{\rm stop}$ prevents the dust from keeping up with the gas flow. However, already for this low $\alpha$ the dust-to-gas ratio gradient is lower than predicted by the purely advective regime. This is the result of turbulent diffusion and acts to reduce the mass flux (as can be seen from \autoref{fig:flux-efficiency} and also the dust-to-gas ratio being below 1 at $z=5H$).

\begin{figure*}
	\includegraphics[width=\textwidth]{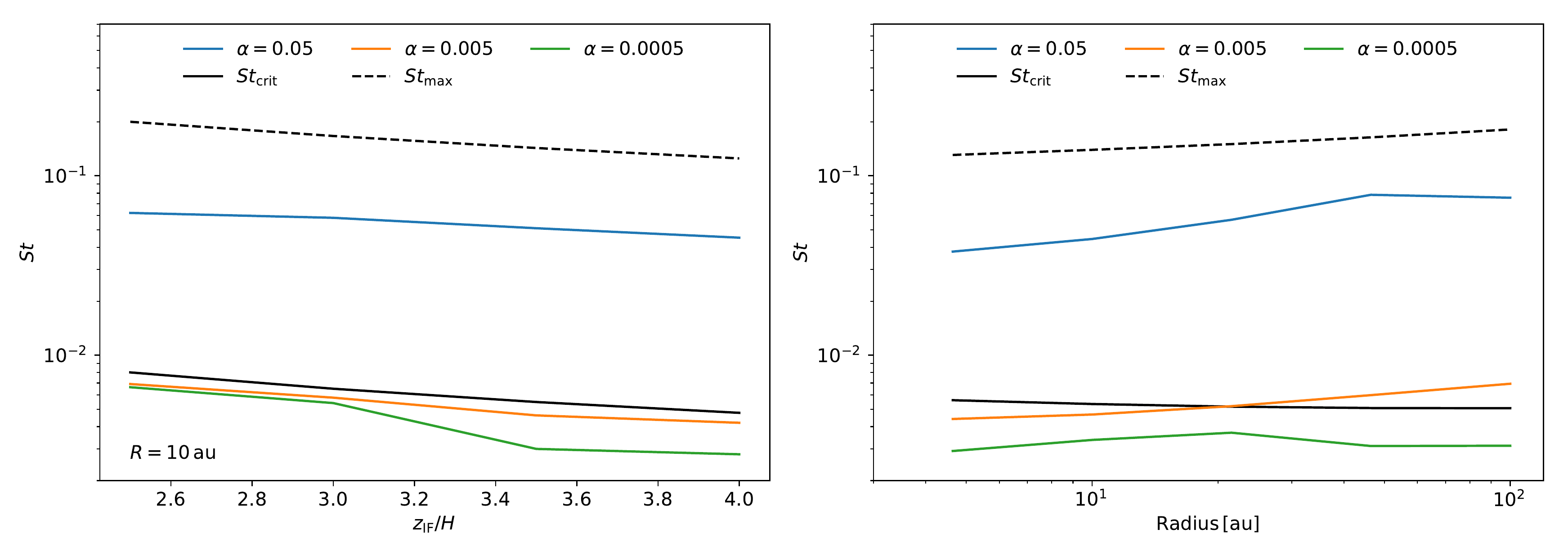}
    \caption{Comparison of the maximum size of dust grains delivered in the wind (defined as the size at which the efficiency drops below 0.5) to analytic estimates of the maximum size of grain delivered to the wind base ($St_{\rm crit}$) and carried away by the wind ($St_{\rm max}$). Left: effect of the ionization front  location at $R=10{\rm au}$. Right: variation with radius for a model with the ionization front height computed using an EUV model with an ionizing flux of $\Phi = 10^{42}\,{\rm s}^{-1}$. In each case the Stokes number shown is the Stokes number in the disc as measured at the ionization front.}
    \label{fig:St_crit}
\end{figure*}

Increasing $\alpha$, one would expect the dust-to-gas ratio to be driven towards a constant value even more strongly. However, in the simulation with $\alpha = 0.05$, we see a negative gradient in dust-to-gas ratio, evidence that diffusion is now driving an outward flux of dust (which is further supported by the average dust velocity in \autoref{fig:profiles} being larger than the mean gas velocity). This gradient is particularly strong close to the ionization front. Our proposed explanation for this is that the diffusive supply of dust to the ionization front from the disc is not matched by a return flux from the wind since the low density downwind of the front means that such particles are not effectively coupled to turbulent motions driving them back through the front.
Conversely, for particles small enough that they remain coupled to the gas through the ionization front, the diffusive flux is cancelled by the $w_{\rm hs}$ correction term resulting in $\epsilon_{\rm F} \approx 1$.

Even in the diffusive regime the mass-loss rate of dust becomes negligible once the Stokes number exceeds $St_{\rm crit}$ by more than a factor of $\sim 10$. For grains of this size, $St > St_{\rm max}$, and the drag force on particles that pass through the ionization front is no longer sufficient to overcome gravity in the wind. Therefore, even if dust can be supplied to the wind, ultimately it can not escape the disc.

The results are not sensitive to the underlying parameters of the disc. This is demonstrated in Appendix \ref{sec:appendix} and \autoref{fig:St_crit}, where we compare the Stokes number at which $\epsilon_{\rm F}$ drops to 0.5 to the values of  $St_{\rm crit}$ and $St_{\rm max}$. We do this varying $z_{\rm IF}$ at $R=10\,{\rm au}$ (left panel) and also for a model where $z_{\rm IF}$ is computed according to \autoref{eqn:rho_ion} for $\Phi = 10^{42}\,{\rm s}^{-1}$. The EUV model follows \citet{HC20}, who assume the density at the wind base is controlled by recombination: 
\begin{equation}
\rho_{\rm ion} = 0.2 m_{\rm H} \left(\frac{3 \Phi}{4 \upi \alpha_2 R^3} \right)^{1/2}, \label{eqn:rho_ion}
\end{equation}
where the velocity at the wind base is $0.5c_{\rm s,wind}$ as before.  Here $m_{\rm H}$ is the mass of a hydrogen atom, $\alpha_2 = 2.6\times 10^{-13}\,{\rm cm^3 s^{-1}}$ is the Case B recombination coefficient and $\Phi$ is the stellar EUV luminosity.

In \autoref{fig:St_crit} we see that the maximum size entrained is close to $St_{\rm crit}$ for $\alpha < St_{\rm crit}$ independent of the height of the ionization front or the location in the disc. Furthermore, although the size increases above $St_{\rm crit}$ for $\alpha > St_{\rm crit}$, it always remains smaller than $St_{\rm max}$.

\subsection{Typical grain sizes entrained}

Now that we have ascertained that the maximum size of dust grains delivered to the wind is determined by $St_{\rm crit}$ in the advective regime while $St_{\rm max}$ limits the size of grains removed in the diffusive regime, we consider what these Stokes numbers mean in terms of grain size. From the definition of these limits (i.e. zero acceleration of a stationary grain just
below and just above the ionization front respectively) 
we can write
\begin{align}
    s_{\rm crit} & =  \sqrt{\frac{8}{\upi}} \frac{\dot{\Sigma}}{\rho_{\rm grain} \Omega} \frac{H_{\rm IF}}{z_{\rm IF}} \left(1 + \frac{z_{\rm IF}^2}{R^2}\right)^{3/2}  \\ 
     & \approx 0.63 
        \left(\frac{\dot{\Sigma}}{10^{-12}\,{\rm g\,cm^{-2}\,s^{-1}}} \right) 
        \left(\frac{R}{10\,{\rm au}}\right)^{3/2} \nonumber \\
        & \qquad \times \; \left(\frac{z_{\rm IF}}{4 H_{\rm IF}}\right)^{-1} \left(\frac{M_*}{1 M_\odot}\right)^{-1/2} \left(\frac{\rho_{\rm grain}}{1\,{\rm g\,cm}^{-3}} \right)^{-1} \micron. \label{eqn:s_crit}
\end{align}
This shows that maximum size of dust particle that can be entrained is insensitive to the disc mass, which only enters through the dependence of $z_{\rm IF}$ on disc mass, which is weak. Note that this equation is valid even if the disc is not vertically isothermal (as assumed in this paper) as long as $H_{\rm IF}$ is determined from $c_{\rm s, disc}$ measured at the ionization front. Similarly the definition for $s_{\rm max}$ follows by replacing $c_{\rm s, disc}$ with $c_{\rm s, wind}$ (in the definition for $H_{\rm IF}$).

\begin{figure}
	\includegraphics[width=\columnwidth]{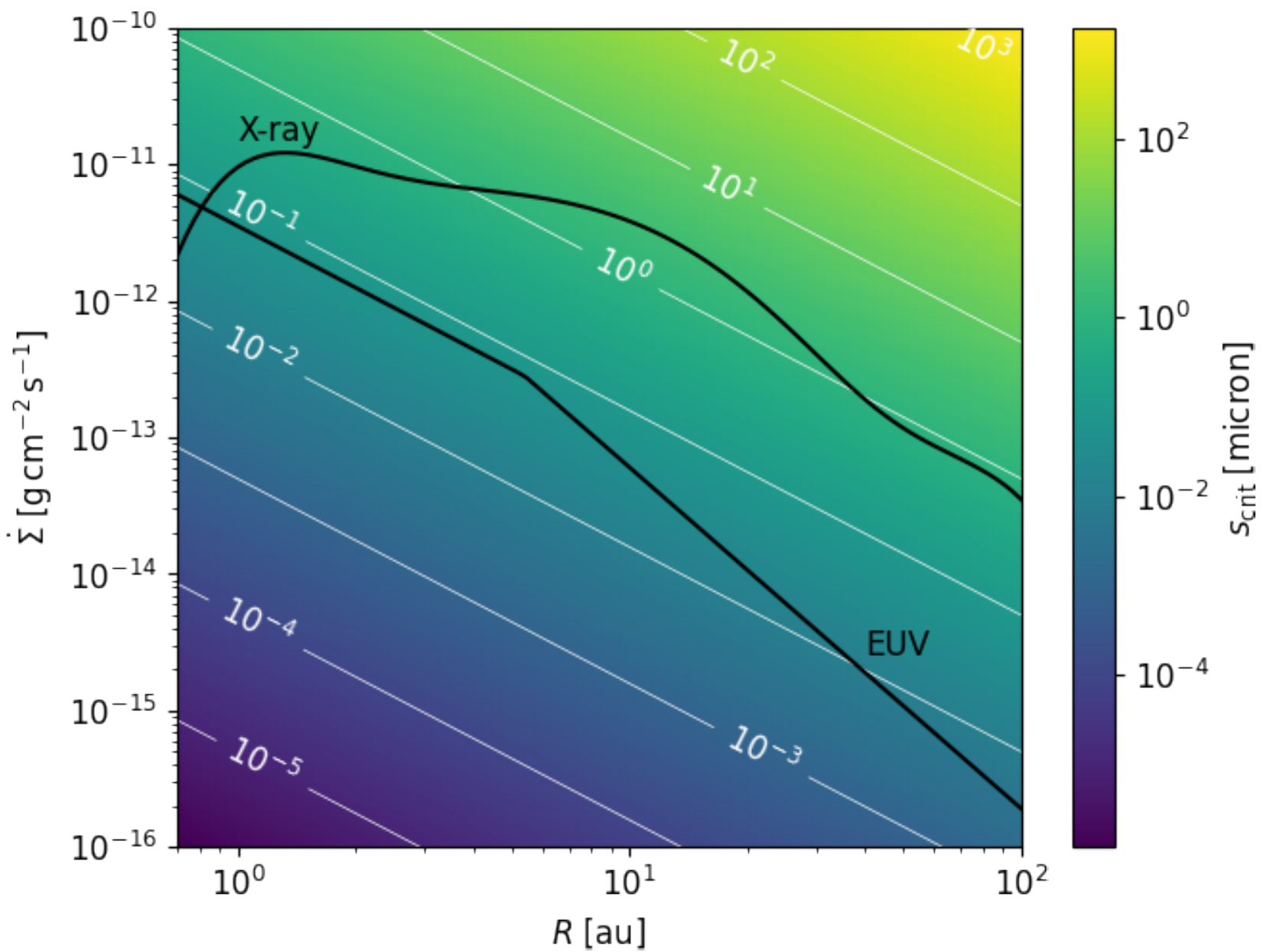}
    \caption{Maximum size of dust particles that can be delivered to a photoevaporative wind for a given mass-loss rate per unit area and radius (colour map and white contours). The black lines show the mass-loss profiles for the EUV driven wind model with $\Phi = 10^{42}\,{\rm s}^{-1}$ \citep{Hollenbach1994} and X-ray driven wind with $L_{\rm X} = 2 \times 10^{30}\,{\rm erg\,s}^{-1}$ \citep{Picogna2019}, for which the integrated mass loss rates are $7.4 \times 10^{-10}$ and $2.7\times 10^{-8}\,M_{\odot}\,{\rm yr}^{-1}$, respectively. Note that $s_{\rm crit}$ is given here for the case $\rho_{\rm grain}=1\,{\rm g\,cm}^{-3}$ and an ionization front height of $z_{\rm IF} = 4 H$ and scales with $H/z_{\rm IF}\rho_{\rm grain}$ (\autoref{eqn:s_crit}).}
    \label{fig:scrit_grid}
\end{figure}

We show $s_{\rm crit}$ for representative values of $\dot{\Sigma}$ and $R$ in \autoref{fig:scrit_grid}, over which we plot the mass-loss profiles from representative EUV \citep{Hollenbach1994} and X-ray \citep{Picogna2019} driven wind models assuming $z_{\rm IF} = 4 \rm H_{\rm IF}$. Typical grain sizes vary between 0.01 and $1\,\micron$.

These values of $s_{\rm crit}$ can be estimated analytically from the mass-loss rates, i.e. in the case of an EUV driven wind with density profile given by (\ref{eqn:rho_ion}) (as is appropriate to disc radii interior to  $R_{\rm g} = GM_*/c_{\rm s,wind}^2 \approx 5\,{\rm au}$):
\begin{align}
    s_{\rm crit,\,EUV} &= 0.2 \left(\frac{6 \Phi}{G M_*  \alpha_2 \upi^2} \right)^{1/2} \frac{m_{\rm H} v_{\rm wind}}{\rho_{\rm grain}} \frac{H_{\rm IF}}{z_{\rm IF}} \label{eqn:s_crit_EUV} \\
        &\approx 0.022  \left(\frac{\Phi}{10^{41}  {\rm s}^{-1}}\right)^{1/2} \left(\frac{M_*}{1 M_\odot}\right)^{-1/2}  \left(\frac{z_{\rm IF}}{4 H_{\rm IF}}\right)^{-1} \micron.
\end{align}
 Note that in the original model of  \citet{Hollenbach1994}, the density at the ionization front
 falls off more steeply with radius beyond $R_{\rm g}$, scaling as  $R^{-2.5}$, outside the gravitational radius:
 this effect has been included in the estimate for $\dot{\Sigma}(R)$ in the EUV case  shown in \autoref{fig:scrit_grid}. 

A simple estimate for $s_{\rm crit}$ in X-ray driven winds may be estimated from ${\dot{\Sigma} \sim 2 \times 10^{-12} (R/10\,{\rm au})^{-3/2}\,{\rm g\,cm^{-2}\,s}^{-1}}$ for an X-ray luminosity, $L_{\rm X} = 2 \times 10^{30}\,{\rm erg\,s}^{-1}$  \citep{Picogna2019}. The corresponding estimate for the maximum grain size entrained is then
\begin{equation}
    s_{\rm crit,\,X-ray} \sim 1 \times \left(\frac{L_{\rm X}}{2 \times 10^{30}\,{\rm erg\,s}^{-1}}\right) \left(\frac{z_{\rm IF}}{4 H_{\rm IF}}\right)^{-1} \micron.
\end{equation}
More precise numbers can obtained by directly using the fits for $\dot{\Sigma}$ provided by \citet{Picogna2019}, which were used in \autoref{fig:scrit_grid}.

Finally, the distribution of dust entrained in the wind can be computed from the flux efficiency, $\epsilon_{\rm F}$ (\autoref{fig:flux-efficiency}). Since $\epsilon_{\rm F}$ falls off rapidly for $St > St_{\rm crit}$ the contribution from sizes much beyond $St_{\rm crit}$ can be neglected.

\section{Discussion}
\label{sec:discussion}
In this paper we have demonstrated that the removal of dust from protoplanetary discs by winds is driven by advection unless turbulence in the disc is strong, i.e. $\alpha \gtrsim St_{\rm crit} \approx 0.01$ (\autoref{eqn:stcrit}). Note that $St_{\rm crit}$ is measured in the disc immediately below the ionization front and thus varies only weakly with system parameters (see \autoref{fig:St_crit}), being mainly set by the difference in temperature between the disc and the ionized wind. We find that strong turbulence acts to increase the amount of dust supplied to the wind by a factor of a few in certain size ranges (see left hand panel of \autoref{fig:flux-efficiency}), rather than decreasing it, as was found by \citet{HC20}. This difference can be attributed to the way in which diffusion was treated in the two studies. \citet{HC20} treated diffusion by adding a diffusive flux to the mass-conservation equation using the model of \citet{Dubrulle1995}. Here we have used a Monte-Carlo model for the dust in which diffusion is treated through the direct coupling of dust to turbulent motions  in the disc gas via drag forces,  modelling the turbulence assuming isotropic Gaussian turbulence with a constant velocity dispersion. The explanation for this difference is that if one simply adds a diffusive flux using the \citet{Dubrulle1995} model, it implies an increase in the dust to gas ratio across the front which can drive a strong negative diffusive flux. The reason why this does not happen in reality is that, as particles cross the ionization front, they decouple from the gas flow and do not participate in diffusive motions. This behaviour can only be captured by a treatment that explicitly models the ability of particles with finite stopping time to decouple from the diffusive motions over a region where there is a steep gradient in background gas properties. We therefore caution against applying the \citet{Dubrulle1995} model in situations involving ionization fronts.

Although isotropic turbulence is likely a poor approximation at the ionization front, our results are unlikely to be substantially affected by this. This is obviously the case when the delivery of dust is dominated by advection, which is the case for both weak turbulence and sufficiently small particles in the regime of strong turbulence. Since large particles cross the ionization front within a stopping time, they cannot couple to the gas within the ionization front, and therefore the details of the turbulence at the ionization are not important. For intermediate grain sizes in conditions of strong turbulence, particles begin to decouple within the ionization front. The mass-loss rate of these particles is sensitive to the width of the ionization front, and therefore possibly also sensitive to details of the turbulence there. However, the dependence of the flux on the properties of the ionization front is weak, and the phenomenological behaviour is unlikely to be affected.

Our results suggest that vertical advective transport in discs could play an important role in determining the vertical height of discs measured in scattered light. Recent non-ideal MHD simulations suggest that discs may have weak turbulence ($\alpha \lesssim 10^{-4}$), with observational studies providing supporting evidence \citep[see, e.g.][]{Mulders2012,Flaherty2015,Simon2018, Flaherty2020}. Under such conditions our models show that advective transport due to the wind should dominate the lofting of small grains -- this could be tested by comparing resolved observations of disc thickness in the scattered light to the thickness derived for millimetre grains \citep[e.g.][]{Pinte2016,Avenhaus2018,Villenave2020}. If turbulence is stratified, i.e. $\alpha$ increases with height, then turbulence might still play an important role. However, \citet{Riols2018} found that the vertical variation of $\alpha$ could not explain the lofting of grains seen in their MHD simulations unless advective transport was also included. We suggest that this is likely to be a general feature of discs undergoing mass loss due to winds, independent of the winds' origin.

In our calculations we have neglected the influence of radiation pressure on the dust grains. Since the optical photosphere is at lower altitudes than the EUV (or X-ray) photosphere, \citet{Owen2019} argued that radiation pressure could remove grains efficiently. We now show that when including advective transport, radiation pressure does not greatly change the picture. Neglecting turbulence, but including radiation pressure, the vertical and radial velocities of dust grains are given by
\begin{align}
    v_{\rm z} &= v_{\rm g} - (1 - \beta) \Omega z St, \\
    v_{\rm R} &= \beta \Omega R St,
\end{align}
where $\beta$ is the ratio of the radiation pressure force to the gravitational force. Here we have assumed that $\beta$ is large enough that the radiation pressure term dominates over all other components (such as the radial gas pressure gradient) in the the equation for $v_{\rm R}$ from \citet{Owen2019}.  Re-writing the gas velocity in terms of $St_{\rm crit}$, we find
\begin{equation}
    \frac{v_{\rm z}}{v_{\rm R}} = \frac{z}{R} \left[ 1 + \frac{1}{\beta}\left(\frac{St_{\rm crit}}{St} -1 \right) \right].
\end{equation} 
If the height of the ionization front scales as $z_{\rm IF} \sim R^{1 + \delta}$, then particles with $St \lesssim St_{\rm crit} / (1 + \beta \delta)$ will be delivered to the wind. For typical values of $\beta$, the maximum grain size delivered to the wind is not much affected. Note that although radiation pressure can increase the maximum size of particles that can be entrained when $\delta < 0$, this is not the case for our EUV model with $\Sigma \propto 1/R$. Another factor that could reduce the size entrained would be if there was a steep dependence of $s_{\rm crit}$ with radius, such that radiation pressure drives the grains to larger radii where they can no longer be entrained. However, given that the dependence of $s_{\rm crit}$ is not particularly strong (\autoref{fig:scrit_grid}), this will also not dramatically affect the maximum size of grains delivered to the winds.

\section{Conclusions}
\label{sec:conclusions}

We have investigated the entrainment of dust grains in photoevaporative winds using a novel Monte-Carlo dust dynamics model which correctly models dust transport across the ionization front separating the disc and ionized wind. This treatment avoids spurious effects previously found when solving the advection-diffusion equation in the limit that the width of the ionization front is less than the dust stopping distance. Our calculations yield dust  transport efficiencies that converge in the limit of narrow ionization fronts as expected. We highlight that special care needs to be taken in the choice of Monte-Carlo dust modeling algorithm in the demanding case of a steep density feature such as an ionization front and that algorithms in the literature produce numerical artefacts  under these conditions. 

Our simulations show that the delivery of dust to the wind base is dominated by the advection of small dust grains by the vertical gas flow that appears as a consequence of the photoevaporative mass loss. This is contrary to the usual assumption that turbulent diffusion is responsible for lofting grains to the ionization front, which we show only occurs if disc turbulence is strong (i.e. for values of the \citealt{Shakura1973} $\alpha$-parameter $\gtrsim 0.01$). 

 Our results confirm the hypothesis of \citet{HC20} that the maximum size of dust grains entering the wind is set by the condition of zero force on a stationary  dust grain immediately below the ionization front (a limit that we denote as $s_{\rm crit}$). This is {\it not} the same as the commonly assumed limit (which, following \citealt{HC20} we designate $s_{\rm max}$) which corresponds to the condition of zero force on a stationary dust grain immediately above the ionization front. The drag force scales as the product of the gas flux and the local sound speed: since the flux is conserved across the front, this means that the ratio of $s_{\rm crit}$ to $s_{\rm max}$ is given by the ratio of local sound speeds, and is thus typically around  $\sim 0.1$  in the case of ionized winds from protostellar discs.  Equation (\ref{eqn:s_crit}) allows the value of $s_{\rm crit}$ to be estimated for any wind where the local mass flux and height of the base of the
 heated region  is known; Figure \ref{fig:scrit_grid} illustrates the typical values that apply in the case of mass loss profiles for canonical EUV and X-ray driven winds. These values are lower by  around a factor $10$, for equivalent parameters, than those previously proposed  \citep{Takeuchi2005,Owen2011a,Franz2020}, a result that we ascribe to the aforementioned difference between $s_{\rm crit}$ and $s_{\rm max}$. 

\section*{Acknowledgements}

We thank Mark Hutchison for many interesting discussions on this topic and James Owen for encouraging us to look into radiation pressure. RAB and CJC acknowledge support from the STFC consolidated grant ST/S000623/1. This project has received funding from the European Research Council (ERC) under the European Union’s Horizon 2020 research and innovation programmes PEVAP (grant agreement No. 853022) and DUSTBUSTERS (grant agreement No 823823). This work was performed using the Cambridge Service for Data Driven Discovery (CSD3), part of which is operated by the University of Cambridge Research Computing on behalf of the STFC DiRAC HPC Facility (www.dirac.ac.uk). The DiRAC component of CSD3 was funded by BEIS capital funding via STFC capital grants ST/P002307/1 and ST/R002452/1 and STFC operations grant ST/R00689X/1. DiRAC is part of the National e-Infrastructure.

\section*{Data Availability}

The simulation code used in this project is available on github at \url{https://github.com/rbooth200/MC_dust} and the simulation results will be shared upon reasonable request.



\begin{figure*}
	\includegraphics[width=\textwidth]{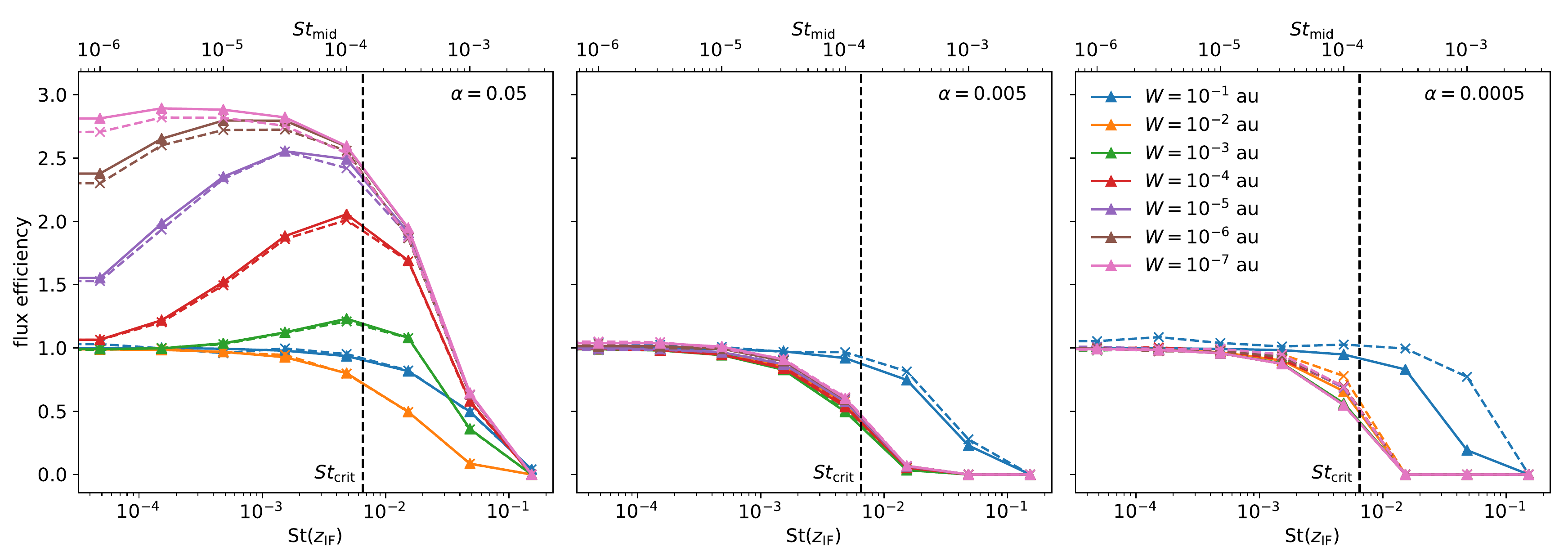}
    \caption{Same as \autoref{fig:flux-efficiency} but for a model with an ionization height $z_{\rm IF}=3H$.}
    \label{fig:flux-efficiency_Z3}

	\includegraphics[width=\textwidth]{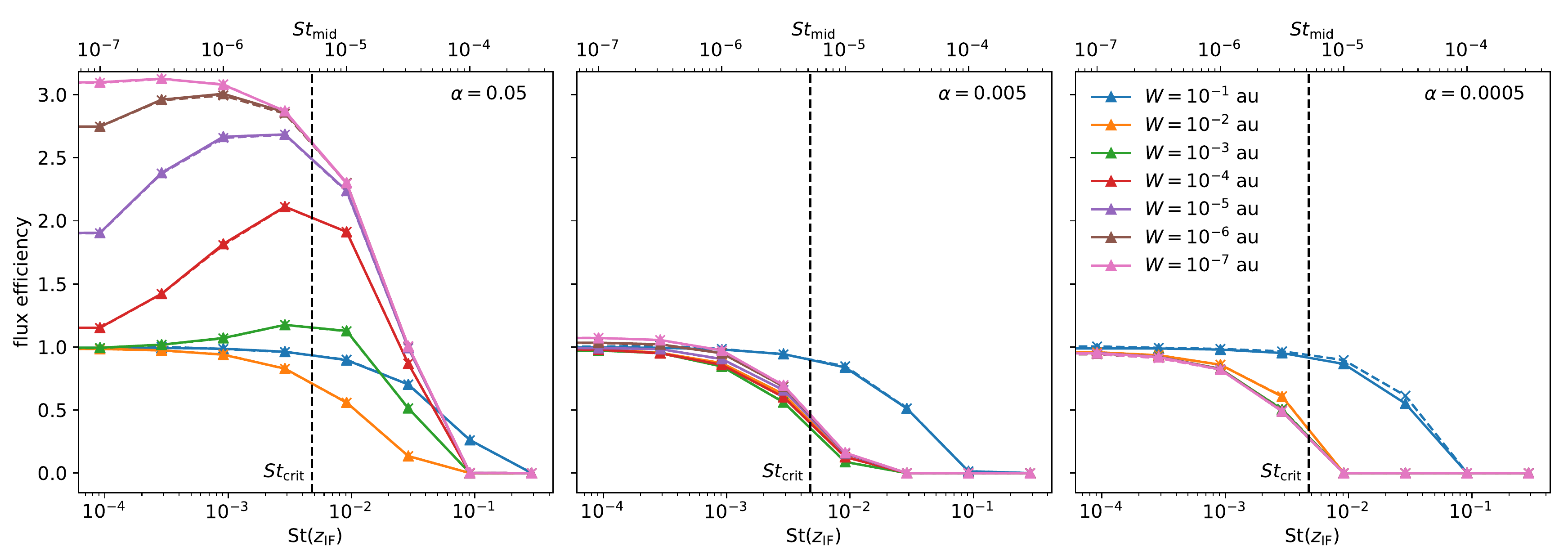}
    \caption{Same as \autoref{fig:flux-efficiency} but for a model with an ionization height $z_{\rm IF}=4H$.}
    \label{fig:flux-efficiency_Z4}
\end{figure*}

\bibliographystyle{mnras}
\bibliography{dusty_winds}



\appendix

\section{Extra flux-efficiency plots}
\label{sec:appendix}
In \autoref{fig:flux-efficiency_Z3} and \autoref{fig:flux-efficiency_Z4} show the flux efficiency, $\epsilon_{\rm F}$, for models with ionization front heights, $z_{\rm IF}$ of $3H$ and $4H$ respectively. The results are nearly identical to case with $z_{\rm IF} =3.5H$ presented in \autoref{sec:results}.


\bsp	
\label{lastpage}
\end{document}